\documentstyle[preprint,aps]{revtex}

\begin{document}
\draft
\title{ \bf Superconducting, magnetic, and charge correlations \\
in the doped two-chain Hubbard model }
\author{Yoshihiro Asai}
\address{Fundamental Physics Section, Electrotechnical Laboratory, \\
Umezono 1-1-4, Tsukuba, Ibaraki 305, Japan}
\date{Received on January 27, 1995}
\maketitle
\begin{abstract}
We have studied the superconducting, magnetic and charge correlation
functions and the spin excitation spectrum in the doped two chain Hubbard
model by projector monte carlo and Lanczos diagonalization methods.
The exponent of the interchain singlet superconducting correlation
function, $\gamma$ , is found to be close to $2.0$ as long as two distinct
non-interacting bands cross the Fermi level.
Magnetic and charge correlation functions decay more rapidly than or as
fast as the interchain singlet superconducting correlation
function along the chains.
The superconducting correlation in the doped two chain Hubbard model
is the most long range correlation studied here.
Implications of the results for the possible universality class of the
doped two chain Hubbard model are discussed.

\end{abstract}
\begin{center}
\bf{I. INTRODUCTION}
\end{center}

In relation to possible superconductivity in the two dimensional
Hubbard model~\cite{ScalapinoR,DagottoR,Bickers,AsaiPRBrc,AsaiPRB},
the two chain Hubbard model and/or the two chain t-J model, which may
be realized in oxygen deficient $Sr_{n-1}Cu_{n+1}O_{2n}$~\cite{Takano}
have attracted much interest recently.
Although superconductivity has not yet been observed in this
material, it may be worthwhile to study these models because they may
make it possible for us to understand dimensional cross-over between
one and two dimensions. Information obtained in them may be of help
in understanding the two dimensional models. The two dimensional
models are more difficult to study than the one dimensional models
and the two chain
models. Though direct studies of the two dimensional
models~\cite{ScalapinoR,DagottoR,Bickers,AsaiPRBrc,AsaiPRB}
are also very important, here we follow the
argument in Refs.~\cite{Rice,Fabrizio,Noack1}
and study the doped two chain Hubbard model, which is defined by the
following Hamiltonian:
\begin{eqnarray}
H =
-t_{\parallel} \sum_{\langle i,j \rangle \sigma}
( c^\dagger_{i \sigma} c_{j \sigma} + H.C. )
+ U \sum_{i} n ^c _{i \uparrow} n ^c _{i \downarrow} \nonumber \\
-t_{\parallel} \sum_{\langle i,j \rangle \sigma}
( d^\dagger_{i \sigma} d_{j \sigma} + H.C. )
+ U \sum_{i} n ^d _{i \uparrow} n ^d _{i \downarrow} \nonumber \\
-t_{\perp} \sum_{i \sigma} ( c^\dagger_{i \sigma} d_{i \sigma} + H.C. ) .
\label{eq:2chainsHubbard}
\end{eqnarray}
We take $t_{\parallel}=1.0$ and adopt it as the unit of energy
hereafter.
Diagonalization studies of the doped two chain models suggest
that the superconducting correlation function may be rather long
range.~\cite{Rice,DagottoYamaji}
The cluster size accessible by the Lanczos method is obviously too small
compared with the superconducting correlation length in these models
to make a convincing conclusion.
The problem should be examined with more sophisticated methods
in computational physics without any approximation and with sufficiently
large system size.

The recently developed density-matrix real space numerical renormalization
group method has been applied to study the long range superconducting
correlation in the doped two chain Hubbard model.~\cite{Noack1}
The authors concluded that there is an enhancement of the superconducting
correlation function over the non-interacting case on up to as large as
$32 \times 2$ system sizes when $t_{\perp}=1.5, U=8.0$ at $N_e/N_s=0.875$
filling, where $N_e$ and $N_s$ are the number of electrons and the number
of sites,
respectively. They also have found persistence of the spin gap after doping
with the same parameter values.

Though their method seems to be quite useful at least in some of one
dimensional spin systems, we take here a different approach to study
some interesting phases of the doped two chain Hubbard model.
For most of this article we adopt the projector
quantum monte carlo method.
While the method is quite useful to our model in the small-U region,
the method currently has difficulty in the large-U region due to
the fermion sign problem.
Our strategy here is to derive reliable results in the
small-U region, which still has not been explored as well as to get
some preliminary information in the large-U region.
The similarity or dissimilarity of the two regions should be of great
interest.
So long as each numerical method has both pros and cons, the author
believes that the problem should be approached with various numerical
techniques.

\begin{center}
\bf{II. STATIC AND DYNAMIC CORRELATION FUNCTIONS}
\end{center}

Here we study singlet superconducting $<O_iO^{\dagger}_j>$, magnetic
$<S^z_iS^z_j>$ and charge $<n_in_j>-<n>_{ave}^2$ correlation functions
and dynamic spin correlation function $S^{zz}(\vec{k},\omega)$ in the
ground state of the doped two chain Hubbard model.
We have employed the projector monte carlo method to study
the interchain singlet superconducting correlation function
$<O_iO^{\dagger}_j>$, where
\begin{eqnarray}
O^{\dagger}_j = \frac{1}{\sqrt{2}}
  ( c^{\dagger}_{j \uparrow} d^{\dagger}_{j \downarrow}
  - c^{\dagger}_{j \downarrow} d^{\dagger}_{j \uparrow} ),
\label{eq:d-wave}
\end{eqnarray}
magnetic $<S^z_iS^z_j>$ and charge $<n_in_j>-<n>_{ave}^2$ correlation
functions, where
\begin{eqnarray}
S^z_i = \frac{1}{2}(n_{i \uparrow}-n_{i \downarrow})
\label{eq:magnetic}
\end{eqnarray}
and
\begin{eqnarray}
n_i = n_{i \uparrow}+n_{i \downarrow}
\label{eq:charge}
\end{eqnarray}
and the Lanczos and the recursion methods
to study $S^{zz}(\vec{k},\omega)$:
\begin{mathletters}
\label{recursion}
\begin{equation}
S^{zz}(\vec{k},\omega ) =
- \frac{1}{N\pi} Im G(\vec{k},E_0 + \omega + i \eta ),
\label{eq:recursion1}
\end{equation}
\begin{equation}
G(\vec{k},x) = \langle \psi_0 | (S^z_{\vec{k}}) ^\dagger (x - H)^{-1}
S^z_{\vec{k}} | \psi_0 \rangle .
\label{eq:recursion2}
\end{equation}
\end{mathletters}
We have adopted periodic boundary condition throughout
this article. $36 \times 2$ and $24 \times 2$ lattices were used for
projector monte carlo calculations. A $6 \times 2$ lattice was used for
Lanczos and recursion calculations. The projecting time $\tau = 10.0$
and $\tau = 1.5$ in units of $t_{\parallel}$ were used for $U=2.0$ and
$U=6.0$, respectively, in the projector monte carlo calculations.
We have adopted the unrestricted Hartree-Fock (UHF) wavefunction as
a trial function of the projector
monte carlo calculation here.
Details of these methods should be looked up in the previous papers
and references cited therein.~\cite{AsaiPRBrc,AsaiPRB}
The two chain Hubbard model with various parameter sets
at $0.833$ filling ($N_e/N_s=0.833$) is studied here.
The singlet superconducting correlation function in the non-interacting
case is given by:
\begin{eqnarray}
<O_iO^{\dagger}_j>_{U=0} =
\frac{1}{8 \pi^2 l^{\gamma}}
[ 2 - cos(2k_f(0)l) - cos(2k_f(\pi)l)],
\label{eq:sprnoint}
\end{eqnarray}
where $k_f(0)=cos ^{-1}(t_{\perp}+\mu)/2$,
$k_f(\pi)=cos^{-1}(t_{\perp}-\mu)/2$ and $\gamma=2$, when $U=0.0$.
The upper bound of $<O_iO^{\dagger}_j>$ in the non-interacting case
is given by $0.05R^{-2}_{ij}$, where $l$ and $R_{ij}$ are distance
between the i-th and j-th sites.

First of all, we study long-range behaviors of some of static
correlation functions in small-U region of the model.
$<O_iO^{\dagger}_j>$ when $t_{\perp}=1.5, U=2.0$ and $t_{\perp}=2.5, U=2.0$
were calculated and they are plotted along with the upper bound of the
non-interacting case in Fig. \ref{fig:sprlnsc}.
It should be noted that while two distinct bands cross the Fermi level
when $t_{\perp}=0.5, U=0.0$ and $t_{\perp}=1.5, U=0.0$,
one of them is raised in energy and does not cross the Fermi level when
$t_{\perp}=2.5, U=0.0$.
The correlation function with $t_{\perp}=2.5, U=2.0$ is strongly suppressed
from the non-interacting one, which does not contradict the result
obtained from the more general superconducting correlation
function.~\cite{AsaiPRBrc}
The suppression may originate from the separation of the bands.
The correlation function with $t_{\perp}=1.5, U=2.0$ is somewhat enhanced or
remains comparable to the non-interacting correlation function.
This parameter region should be examined more carefully.

The superconducting correlation function $<O_iO^{\dagger}_j>$
with $t_{\perp}=1.5, U=2.0$ was re-plotted along with the upper bound of
the non-interacting case on a log-log scale in Fig. \ref{fig:sprU2tv15}.
The correlation function has inversion symmetry at the inversion center:
$|i-j|=17.5$. We show only the left hand side of the function.
Though the correlation function itself is somewhat enhanced over the
non-interacting one, the exponent of the correlation
function $\gamma$ remains close to that of the non-interacting case:
$2.0$.

At this stage, we compare the long range behaviors of the interchain
singlet superconducting, magnetic and charge correlation functions.
We have calculated these correlation functions with $t_{\perp}=1.5, U=2.0$
in $24 \times 2$ lattice and plotted them in Fig. \ref{fig:sprspnchg}.
As mentioned earlier, $<O_iO^{\dagger}_j>$ scales as $0.05R^{-2}_{ij}$.
The data points of the magnetic $<S^z_iS^z_j>$ and
charge $<n_in_j>-<n>_{ave}^2$
correlation functions are found between $\pm 0.05R^{-2}_{ij}$ and that
means the magnetic and the charge correlation functions decay more rapidly
than or as fast as the interchain singlet superconducting correlation
function. The interchain singlet superconducting correlation function is
the most long range correlation function among these three
correlation functions.
This is quite different from finite-U one dimensional Hubbard model.

The superconducting correlation function $<O_iO^{\dagger}_j>$ with
$t_{\perp}=1.5, U=6.0$ and $t_{\perp}=0.5, U=2.0$ were plotted along with
the upper bound of the non-interacting case in Fig. \ref{fig:sprU6tv15}.
$<O_iO^{\dagger}_j>$ for both the two parameter sets scale
as $0.05R^{-2}_{ij}$ like $<O_iO^{\dagger}_j>$
for $t_{\perp}=1.5, U=2.0$ does.
The oscillatory part for $t_{\perp}=1.5, U=6.0$ is more
enhanced than that for $t_{\perp}=1.5, U=2.0$. The oscillatory part
with $t_{\perp}=0.5, U=2.0$ is largely reduced compared with that of
$t_{\perp}=1.5, U=2.0$.
The fermion sign problem in our present version of projector monte
carlo method does not allow us to use larger projecting time than
$\tau = 1.5$ when $t_{\perp}=1.5, U=6.0$ at $0.833$ filling.
The result with $U=6.0$ should be taken preliminary.
Whether or not the large-U region and small-U region can be classified
into the same
universality class (possibly with a spin gap) or not may be a subject
of future interest.

We are not able to decide if the spin gap persists after doping or
not from our data on the correlation function because of the size of the
error bars and
the oscillatory behavior.
The authors of Refs.~\cite{Rice,Noack1,Poilblanc} believe that
the spin gap persists
based on their numerical results. If the spin gap persists after
doping, one of the possibilities is that our model is classified into
the Luther-Emery (LE) liquid phase with exponential decay of the $2k_F$
magnetic correlation function. (see Table II of Ref.~\cite{Shiba})
As the exponent of power-law decay of $<O_iO^{\dagger}_j>$ , $\gamma$ ,
is estimated to be about 2, the exponent of $2k_F$ charge
correlation function
should be about $0.5$ in the LE liquid phase.
This leads to a conclusion that the $2k_F$ charge correlation is the
most long-range. Clearly, our data with $U=2.0$
(Fig. \ref{fig:sprspnchg}) is not consistent with this LE picture.
Some other exotic possibilities consistent with the spin gap should be
pursued.~\cite{Poilblanc}
In spite of slow convergence of $\ln b(2k_F, L/2)$ versus $\ln L$
in the Fig.5 of the previous communication~\cite{AsaiPRBrc}, we may
not be able to exclude a possibility of the Tomonaga-Luttinger (TL)
liquid phase in the small-U region, however.

The dynamic spin correlation function was calculated.
The elementary spin excitation spectrum was obtained by tracing
out $\omega$ , giving the first peak of $S^{zz}(\vec{k},\omega)$, where
$\vec{k}=(k_{\parallel},k_{\perp})$ with $k_{\parallel}$ defined along
the chains and $k_{\perp}$ defined vertical to the chains.
It should be noted that the spectrum can be also obtained from the
energy eigenvalues in subspaces with definite $\vec{k}$ values.
Both the $k_{\perp}=0$ and $k_{\perp}=\pi$ branches of the spectra
are plotted in Figs. \ref{fig:spctr}.
Three different parameter sets; $t_{\perp}=0.5, U=2.0$,
$t_{\perp}=1.5, U=2.0$ and $t_{\perp}=1.5, U=8.0$ have been used.
The spectra at half-filling ($N_e/N_s=1.0$) are shown in the
insets.
Spectra of the three parameter sets at half-filling are
isomorphic to each other irrespective of values of the parameters
in both the $k_{\perp}=0$ and the $k_{\perp}=\pi$ channels.
In the $k_{\perp}=0$ channel, $k_{\perp}=\pi$ is no longer one of the
lowest energy states unlike in the one dimensional half-filled Hubbard
model. At $0.833$ filling, spectra for $t_{\perp}=1.5, U=2.0$
and $t_{\perp}=1.5, U=8.0$ are isomorphic to each other in both
the $k_{\perp}=0$ and the $k_{\perp}=\pi$ channels. The spectrum for
$t_{\perp}=0.5, U=2.0$ is quite different from them.
The spin dynamics in the doped two chain Hubbard model changes with
$t_{\perp}$. The change occurs between $t_{\perp}=0.5$ and
$t_{\perp}=1.5$. We do not observe the change in the half-filled state.
The reduction of the oscillatory part of $<O_iO^{\dagger}_j>$ for
$t_{\perp}=0.5, U=2.0$ may originate from the change in
spin dynamics. The change has no effect on $\gamma$.
In the non-interacting case, we have two separated bands when
$t_{\perp} \neq 0$. Both of the two bands are partially filled
in both the two cases studied here; $t_{\perp} = 0.5$ and
$t_{\perp} = 1.5$.
It seems unlikely that the change in the spin dynamics is related
with the number of partially filled bands.
The Fermi wave vector of the second band (higher in energy)
$k_{F2}$ in the noninteracting case is $\pi/3$, which is closest
to $\pi/2$ in this system size.
The spin dynamics when $t_{\perp} = 0.5$ and $U = 2.0$ at $0.833$
filling may be related with the phase with two gapless spin modes
and one gapless charge mode found by using
weak coupling renormalization group
techniques (the C1S2 phase in the notation of the authors
of Ref.~\cite{Balents}).

It should be noted that the smallest energy differences with the
ground state are given at $(k_{\parallel},k_{\perp})=(0.17\pi,0.0),
(0.50\pi,0.0),(0.67\pi,0.0)$ and $(0.83\pi,0.0)$ for
$t_{\perp}=0.5, U=2.0$ and at $(0.17\pi,\pi),(0.50\pi,\pi),(0.67\pi,\pi)$
and $(0.83\pi,\pi)$ for $t_{\perp}=1.5,U=2.0$ and $t_{\perp}=1.5,U=8.0$,
respectively. The smallest energy differences for these parameter sets
are $0.08, 0.10$ and $0.37$, respectively.
The possible spin gap for $U=2.0$ is less clear than that with $U=8.0$
in this small system size.

\begin{center}
\bf{III. CONCLUSIONS}
\end{center}

We have calculated the interchain singlet superconducting, magnetic and
charge correlation functions and the spin excitation spectrum in the doped
two chain Hubbard model.
The exponent of the interchain singlet superconducting correlation
function, $\gamma$ , is found to be close to $2.0$ as long as two distinct
non-interacting bands cross the Fermi level. Our data is not consistent
with the LE picture.
Some other exotic possibilities consistent with the spin gap should be
pursued.
Of the three correlation functions, the interchain singlet superconducting,
magnetic and charge correlation functions, the interchain singlet
superconducting correlation function is the most long range
correlation function in the doped two chain Hubbard model.
This is quite different from finite-U one dimensional Hubbard model.

\begin{center}
\bf{ACKNOWLEDGMENTS}
\end{center}

The author is thankful to the Research Information Processing System
(RIPS) of the Agency of Industrial Science and Technology (AIST) Japan
and the computer center of Institute of Molecular Science (IMS) Japan
for allocating CPU time to the author.
He is also thankful to the referee for leading author's attention
to Ref.~\cite{Balents}, which appeared in the Los Alamos
preprint database server after the submission of the original manuscript.


%
%
\begin{figure}
\caption{The superconducting correlation function $<O_iO^{\dagger}_j>$
at $0.833$ filling on a $36 \times 2$ lattice is plotted versus $|i-j|$.
Circles and squares are the data obtained with $t_{\perp}=1.5, U=2.0$ and
$t_{\perp}=2.5, U=2.0$, respectively.
The upper bound of $<O_iO^{\dagger}_j>_{U=0}$: $0.05R^{-2}_{ij}$ is plotted
as well, where $R_{ij}$ is distance between sites in the rung.}
\label{fig:sprlnsc}
\end{figure}
\begin{figure}
\caption{The superconducting correlation function $<O_iO^{\dagger}_j>$
at $0.833$ filling on a $36 \times 2$ lattice is plotted versus $|i-j|$
on a log-log scale. Closed circles connected with real line are the data
obtained with $t_{\perp}=1.5, U=2.0$.
A dotted line is upper bound of $<O_iO^{\dagger}_j>_{U=0}$: $0.05|i-j|^{-2}$.
The correlation function has inversion symmetry at the inversion
center: $|i-j|=17.5$. We show only the left half of the function.}
\label{fig:sprU2tv15}
\end{figure}
\begin{figure}
\caption{The superconducting $<O_iO^{\dagger}_j>$, magnetic
$<S^z_iS^z_j>$ and charge $<n_in_j>-<n>_{ave}^2$ correlation functions
at $0.833$ filling on a $24 \times 2$ lattice with $t_{\perp}=1.5, U=2.0$
are plotted versus $|i-j|$.
Circles, squares and crosses are the superconducting, magnetic
and charge correlation functions, respectively. The functions
$\pm 0.05R^{-2}_{ij}$ are also plotted as guides.}
\label{fig:sprspnchg}
\end{figure}
\begin{figure}
\caption{The superconducting correlation function $<O_iO^{\dagger}_j>$
at $0.833$ filling on a $36 \times 2$ lattice is plotted versus $|i-j|$.
Circles and squares are the data obtained with $t_{\perp}=1.5, U=6.0$ and
$t_{\perp}=0.5, U=2.0$, respectively.
The upper bound of $<O_iO^{\dagger}_j>_{U=0}$: $0.05R^{-2}_{ij}$ is plotted
as well, where $R_{ij}$ is the distance between sites in the rung.}
\label{fig:sprU6tv15}
\end{figure}
\begin{figure}
\caption{The spin excitation spectrum of a $6 \times 2$ two chain
Hubbard model
at $0.833$ filling. The $k_{\perp}=0$ and $k_{\perp}=\pi$ channel are
plotted in (a) and (b), respectively. Open squares,
closed squares and closed
circles are the spectra obtained with $t_{\perp}=0.5, U=2.0$,
$t_{\perp}=1.5, U=2.0$, and $t_{\perp}=1.5, U=8.0$, respectively.
The horizontal axis is the wavenumber in units of $\pi$.
The corresponding spectra at half-filling ($N_e/N_s=1.0$) are depicted
in the insets.}
\label{fig:spctr}
\end{figure}

\end{document}